\documentclass[aps,prb,twocolumn,superscriptaddress,floatfix,showpacs]{revtex4}
\usepackage{graphicx}
\begin{document}

\title{Magnetic Transformations in the Organic Conductor $\kappa$-(BETS)$_2$Mn[N(CN)$_2$]$_3$
at the Metal-Insulator Transition}

\author{O.~M.~Vyaselev}
\affiliation{Institute of Solid State Physics, Russian Academy of Sciences,
Chernogolovka, Moscow region, 142432, Russia}

\author{M.~V.~Kartsovnik}
\author{W.~Biberacher}
\affiliation{Walther-Meissner-Institut, Bayerische Akademie der Wissenschaften, Garching,
Germany}

\author{L.~V.~Zorina}
\affiliation{Institute of Solid State Physics, Russian Academy of Sciences,
Chernogolovka, Moscow region, 142432, Russia}

\author{N.~D.~Kushch}
\author{E.~B.~Yagubskii}
\affiliation{Institute of Problems of Chemical Physics, Russian Academy of Sciences,
Chernogolovka, Moscow region, 142432, Russia}

\date{\today}

\begin{abstract}

A complex study of magnetic properties including dc magnetization, $^1$H NMR and magnetic
torque measurements has been performed for the organic conductor
$\kappa$-(BETS)$_2$Mn[N(CN)$_2$]$_3$ which undergoes a metal-insulator transition at
$T_{MI}\approx 25\,$K. NMR and the magnetization data indicate a transition in the
manganese subsystem from paramagnetic to a frozen state at $T_{MI}$, which is, however,
not a simple N\'{e}el type order. Further, a magnetic field induced transition resembling
a spin flop has been detected in the torque measurements at temperatures below $T_{MI}$.
This transition is most likely related to the spins of $\pi$ electrons localized on the
organic molecules BETS and coupled with the manganese $3d$ spins via exchange
interaction.

\end{abstract}

\pacs{74.70.Kn, 71.30.+h, 71.27.+a, 75.30.-m, 75.30.Gw, 75.30.Cr, 76.60.Jx}

\maketitle

\section{Introduction}

Quasi-two-dimensional organic charge transfer complexes can be visualized as sheets of
organic donor/acceptor molecules sandwiched between insulating anion/cation
layers.\cite{ishi88} Conductivity in such materials is associated with organic layers.
For years they have been in focus of extensive research activities because these
high-purity materials with relatively simple Fermi surfaces offer rich
pressure-temperature-field ($P$--$T$--$B$) phase diagrams. As a result of the interplay
between electron correlations, the electron-phonon interaction, the electron kinetic
energy, and characteristics of the Fermi surface topology, the phase diagram can include
metal-insulator (MI) and superconducting (SC) transitions as well as different kinds of
charge and spin ordering: charge and spin density waves, long-range antiferromagnetic
(AF) order, spin glass etc. \cite{ishi88,lebe08} Synthesis of radical cation salts of
organic $\pi$-donors with paramagnetic metal complex anions
\cite{r6-1,r6-4,r6-5,r6-6,r6-7} has added a new dimension to the physics of organic
conductors due to the implications of the interaction between the conduction electrons of
the $\pi$ band with localized $d$ electrons. For example, interaction between localized
spins in insulating magnetic layers and itinerant spins in conducting organic layers was
found to lead to new fascinating phenomena such as field-induced superconductivity
observed in $\lambda$-(BETS)$_2$FeCl$_4$ \cite{r10} and
$\kappa$-(BETS)$_2$FeBr$_4$,\cite{r11} where BETS stands for C$_{10}$S$_4$Se$_4$H$_8$,
bis(ethylenedithio)tetraselenafulvalene.

The recently synthesized layered conductor $\kappa$-(BETS)$_2$Mn[N(CN)$_2$]$_3$
\cite{r12} is expected to give a thrilling combination of potentially non-trivial
magnetic properties arising from the nearly triangular network of Mn$^{2+}$ ions in the
anion layer, with strong electron correlations characteristic of the narrow half-filled
conducting band of organic layers. At ambient pressure, this material undergoes a MI
transition at $T_{MI}\simeq 25\,$K, which however can be suppressed by applying a
relatively low external pressure, giving way to a superconducting state with maximum $T_c
= 5.75\,$K at $P = (0.6-1.0)$\,kbar.\cite{ZverevPT} Resistivity measurements combined
with X-ray studies and electronic band structure calculations \cite{ZverevPT} have
suggested the electronic ground state to be a Mott insulator. However, the question about
the role of the interaction between itinerant spins in the donor layers and localized
spins of Mn$^{2+}$ in the formation of the insulating ground state is still open. For
example, it was believed that in $\lambda$-(BETS)$_2$FeCl$_4$ where the MI and the AF
transitions coexist,\cite{BrossFe} the magnetic ordering in Fe$^{3+}$ subsystem leads to
localization of $\pi$-electrons of BETS.\cite{hott00} However, recent specific heat
\cite{AkibaSH} and M\"{o}ssbauer \cite{BrooksMoess} studies have casted doubt on this
viewpoint, suggesting that the magnetic order only exists among the localized
$\pi$-electrons while the Fe$^{3+}$ moments stay paramagnetic.

To clarify the issue of the $\pi-d$ interaction, its influence on the phase diagram of
$\kappa$-(BETS)$_2$Mn[N(CN)$_2$]$_3$, and to understand the driving force of the MI
transition, we have studied its magnetic properties revealed by dc magnetization,
magnetic torque, and NMR measurements. In this paper we report results of this
investigation, arguing for a magnetic transition caused apparently by the MI transition.

\section{Experimental}

The crystal structure of $\kappa$-(BETS)$_2$Mn[N(CN)$_2$]$_3$ is monoclinic with the
space group $P$2$_1$/$c$ and the lattice constants at 90~K \textit{a}=19.453(5)\AA,
\textit{b}=8.381(5)\AA, \textit{c}=11.891(5)\AA, $\beta$=92.784(5)$^\circ$, and
\textit{V}=1936.37\AA$^3$, with two formula units per unit cell \cite{r12}. The
conducting layers are formed by BETS dimers in the (\textit{bc}) plane and sandwiched
between the polymeric Mn[N(CN)$_2$]$_3$ anion layers in the \textit{a} direction. The
crystal growth procedure and details of the structure have been described
elsewhere.\cite{r12, ZverevPT}

The samples were thin-plate single crystals of $\sim 1\times 0.5\times 0.1$\,mm$^3$ size,
with the largest dimension along the conducting BETS layers [crystallographic
(\textit{bc}) plane]. Crystallographic orientations for each crystal were X-ray defined.
The dc magnetization of a 90$\,\mu$g sample was measured  using Quantum Design MPMS-XL
SQUID magnetometer in fields from 1 to 70~kOe. The sample was fixed between two
$6\times6\times1\,$mm$^3$ SrTiO$_3$ (STO) plates for $H\perp (bc)$ and on a
$5\times5\times0.2\,$mm$^3$ Si plate for $H\parallel (bc)$ measurements. Magnetic moments
of the substrates were measured separately for reference. $^1$H NMR was measured
using Bruker MSL-300 spectrometer in fields 14 and 70~kOe on a $\sim100\,\mu$g crystal
attached to a quartz holder with a touch of silicon grease. The empty holder with the
grease gave no $^1$H NMR signal on the scale of the signal from the sample. Magnetic
torque from a 40$\,\mu$g sample was measured in fields up to 150~kOe with a home-made
cantilever beam torque meter described in Ref.~\onlinecite{chri94}. The cantilever was
made of as-rolled beryllium-copper foil 50\,$\mu$m thick.

\section{Results and discussion}

\subsection{dc magnetization}

\begin{figure}[h]
\includegraphics[width=0.65\linewidth,angle=90]{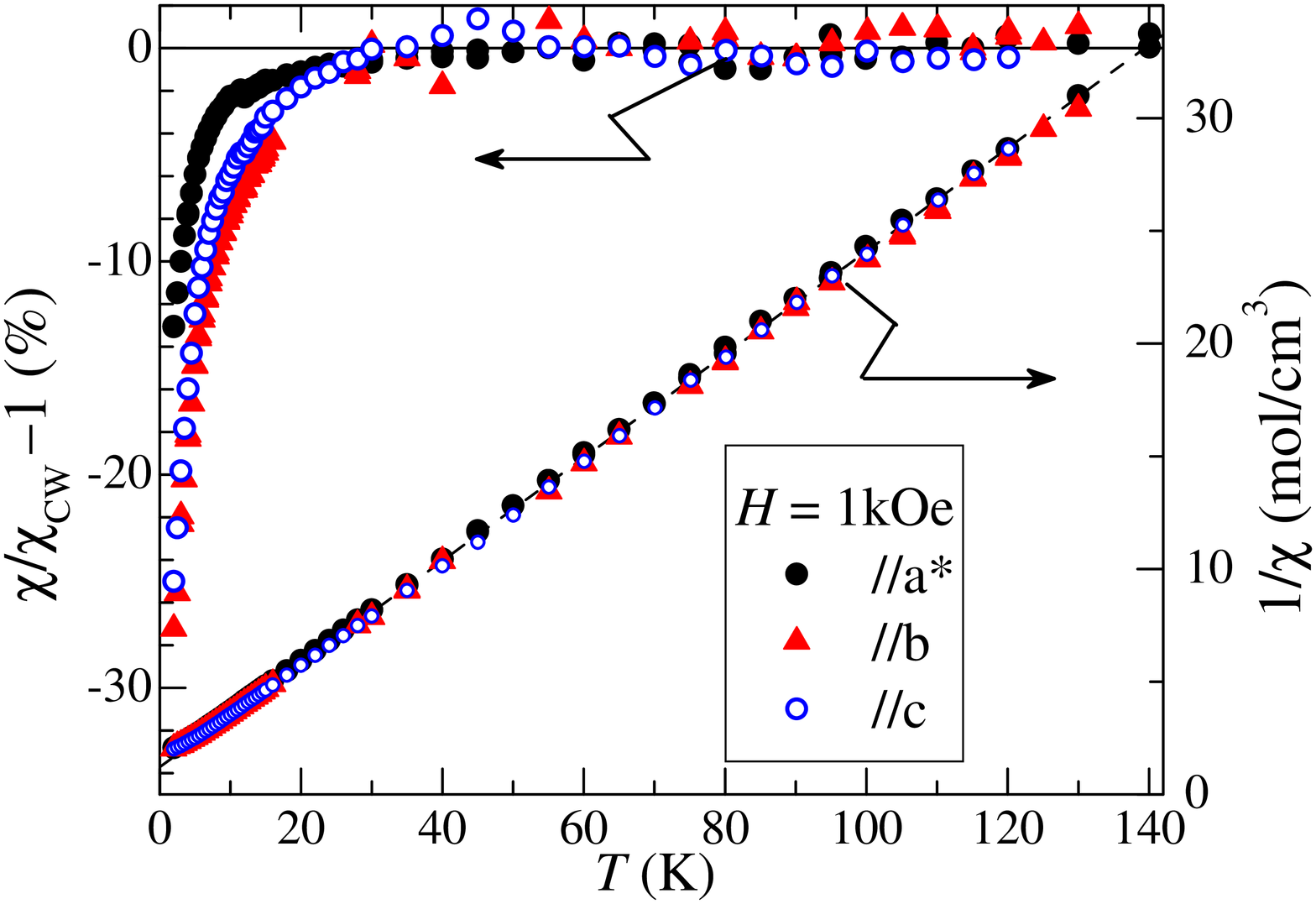}
\caption{(Color online) Temperature dependence of the inverse susceptibility,
$\chi^{-1}=H/M$ (right scale) and the relative deviation of $\chi$ from the Curie-Weiss
value (left scale) at a field of 1~kOe parallel to the $a^{\ast}$-, $b$-, and
$c$-axes. \label{figMvsT}}
\end{figure}

In agreement with the earlier report,\cite{r12} the measured magnetic moment of the
sample grows with diminishing temperature obeying the Curie law down to $T_{\mathrm
MI}\simeq25\,$K: the sample magnetization per mole, $M$, is quite precisely described as
$M=\chi_{CW}H$, where $\chi_{CW}=C_m/(T-\theta)$ is the Curie-Weiss susceptibility. With
the Curie constant $C_m=4.38\,$cm$^3$K/mol calculated for the Land\'{e} factor $g=2$ and
the total angular momentum $J=5/2$, the fits to the experimental data give $\theta=-5.9$,
$-4.6$, and $-4.9\,$K for the field directions along the $a^\ast[\perp(bc)]$, \textit{b},
and \textit{c} axes, respectively. Figure \ref{figMvsT} shows temperature dependences of
the inverse susceptibility, $\chi^{-1}=H/M$, as well as $\chi/\chi_{CW}-1$, the relative
deviation of $\chi$ from the Curie-Weiss value, for the field $H=1\,$kOe applied along
the $a^\ast$, \textit{b}, and \textit{c} directions. One can see that at $T>25\,$K the
susceptibility is essentially isotropic, following accurately the Curie-Weiss law. This
denotes that the system magnetization in this region is dominated by localized moments of
Mn$^{2+}$ in high-spin state ($L=0, S=5/2$) with AF interactions, as follows from the
negative sign of $\theta$.

As the system enters the insulating state below 25~K, the susceptibility turns down from
$\chi_{CW}$. Hardly visible in the $\chi^{-1}(T)$ plot, this deviation is clearly seen in
the $T$-dependence of $\chi/\chi_{CW}-1$, Fig.\ref{figMvsT}. The absolute value of the
drop of $\chi$ from $\chi_{CW}$ at 2~K is $0.9-1.8\times10^{-1}\,$cm$^3$/mol, depending
on the field direction, which is huge compared to the usual value
($\leq10^{-3}\,$cm$^3$/mol) of the conduction electron spin susceptibility known for akin
conducting non-magnetic organic compounds. Therefore the change of the magnetization
behavior below the MI transition temperature should be attributed to a magnetic
transformation in the Mn$^{2+}$ network.

Another evidence for this can be found in the field dependences of the magnetization.
Within the molecular field theory, the exchange interaction between the localized
moments is modeled as an additional magnetic field component $H'=\lambda M$, where
$\lambda$ is related to the Curie-Weiss temperature $\theta$ as $\lambda=C_m/\theta$.
The magnetization, $M$, in the paramagnetic state is then expressed in terms of ($B_{eff}/T$),
where the effective field $B_{eff}=H+\lambda M$.

\begin{figure}[h]
\includegraphics[width=0.65\linewidth,angle=90]{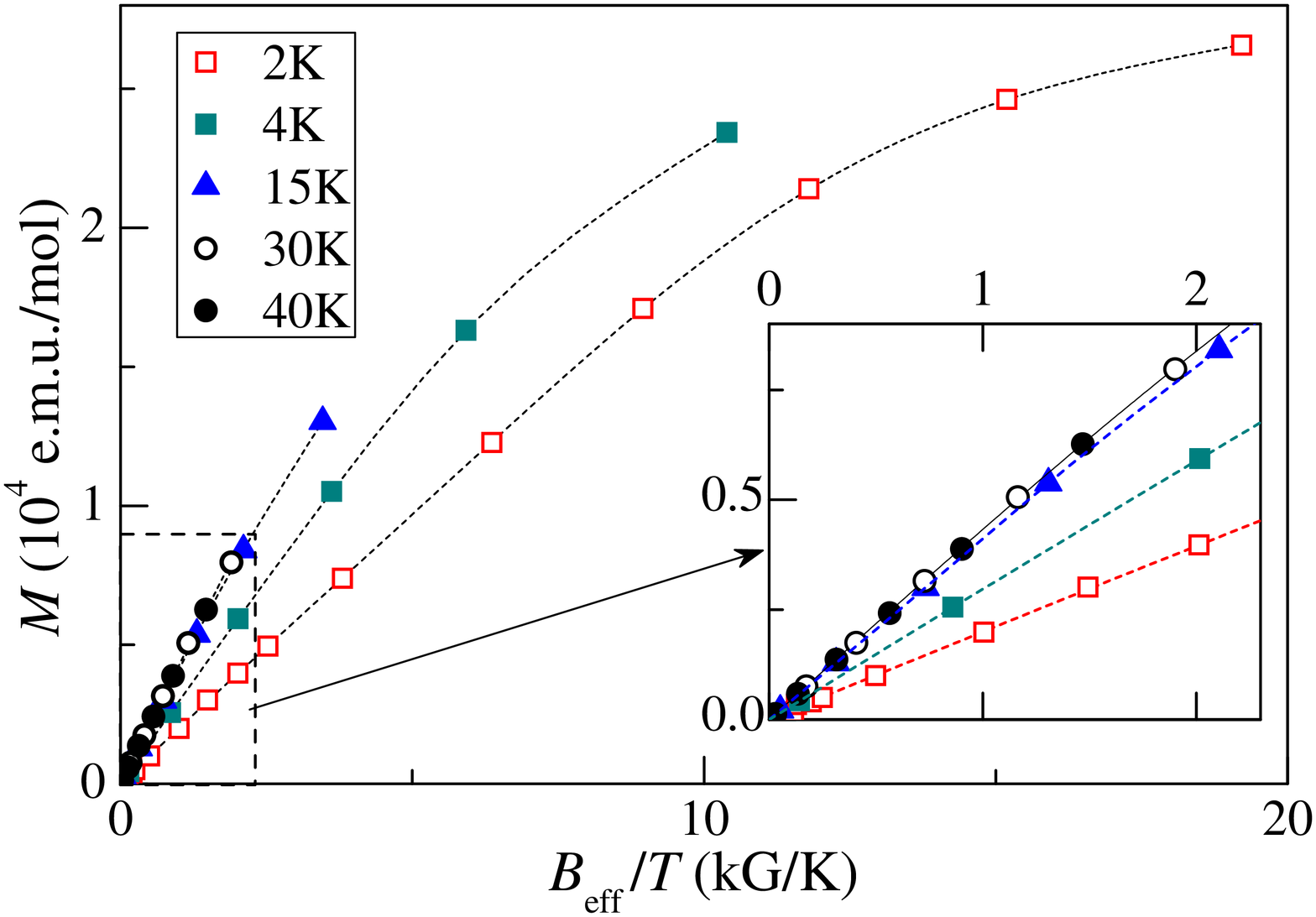}
\caption{(Color online) $c$-axis magnetization versus $B_{eff}/T=(H+\lambda M)/T$
for several temperatures above and below the MI transition, with $\lambda=-0.9$
corresponding to $\theta=-4.9\,$K obtained from the $T$-dependences of the magnetization.
The inset is a close-up to demonstrate the universality of the plot for $T>T_{MI}$.
Dashed lines are guides to the eye. \label{figMvsH}}
\end{figure}

Fig.~\ref{figMvsH} demonstrates the $c$-axis magnetization plotted in function of
$B_{eff}/T=(H+\lambda M)/T$, with $\lambda=-0.9$ corresponding to $\theta=-4.9\,$K
obtained from the $T$-dependences of the magnetization as described above.
The same qualitative behavior has been obtained for the field applied along
the $a^{\ast}$- and $b$-axes.
At temperatures above 25~K, the magnetization is a universal function of ($B_{eff}/T$), as is
more clearly seen in the inset in Fig.~\ref{figMvsH}, which is consistent with the
paramagnetic state of AF-interacting Mn$^{2+}$ ions. At lower temperature the magnetization
declines from the common high-$T$ behavior, falling off more rapidly with diminishing temperature.
This suggests that the state of the Mn$^{2+}$ spin system below $T_{MI}$ is no more
paramagnetic.

\subsection{$^1$H NMR}
The title compound includes 8 inequivalent hydrogen sites which belong to the ethylene
groups, located at the terminals of BETS molecules. Our study of the behavior of $^1$H
NMR spectrum with rotating the magnetic field in the ($ac$) plane at $T=74\,$K has
revealed 8 respective resonance peaks.\cite{NMR} The angular dependence of each peak is
\textit{quantitatively} reproduced by a straightforward calculation of the dipolar field
from the 3$d$ Mn$^{2+}$ ion electron spin moments ($S=5/2, g=2$) (a detailed analysis of
the $^1$H NMR data will be published separately). Furthermore, the proton NMR peak
positions for $H\parallel a^\ast$ at temperatures from 4 to 150~K are linear in dc
magnetization for the same temperatures. This infers that protons probe the dipolar field
induced by Mn$^{2+}$ ions at hydrogen sites.

\begin{figure}[h]
\includegraphics[width=0.65\linewidth,angle=90]{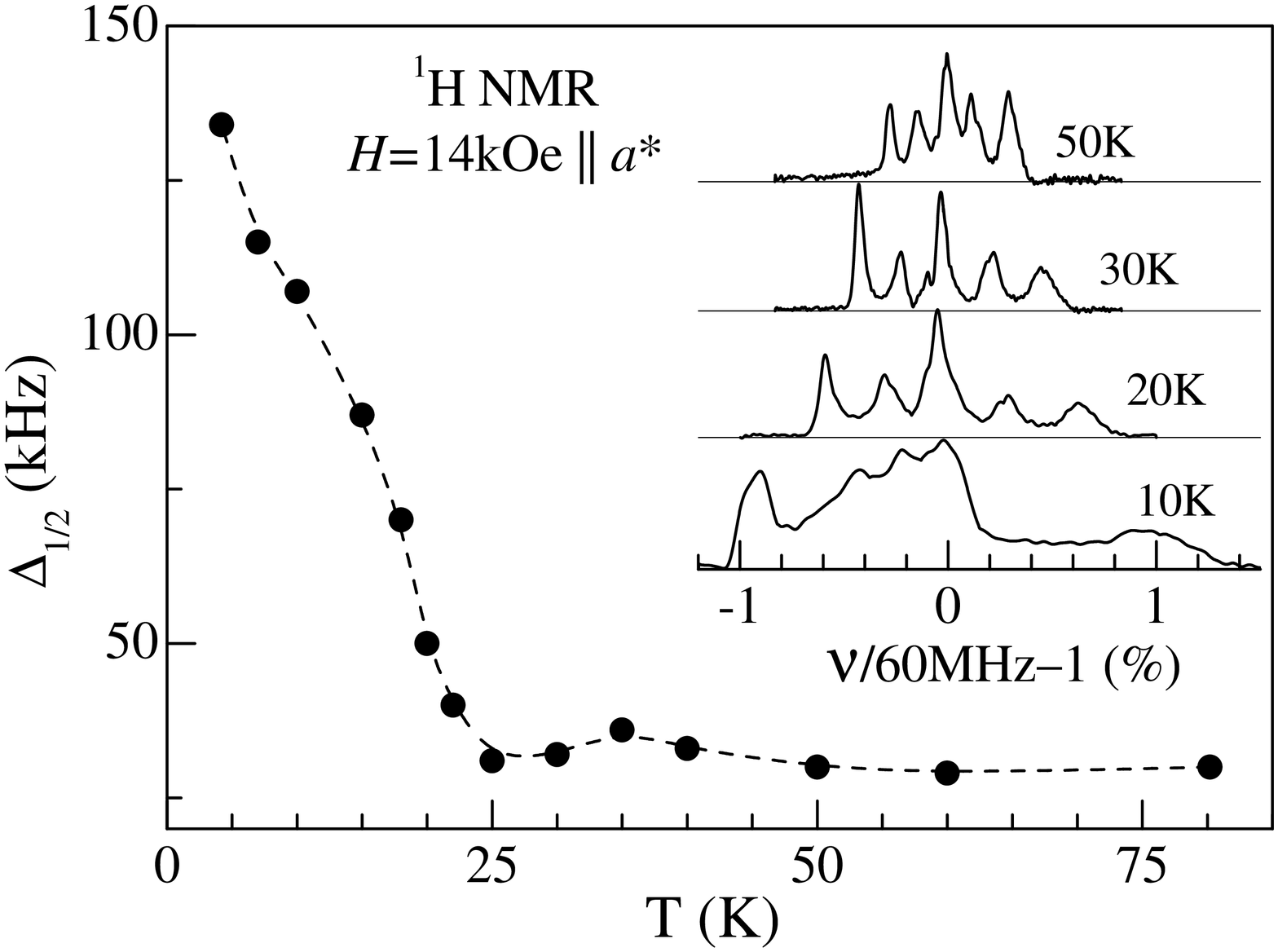}
\caption{Temperature dependence of the linewidth of the leftmost peak in the $^1$H NMR
spectrum obtained at $H=14$\,kOe parallel to $a^\ast$. Inset: NMR spectra at the same
field orientation, at different temperatures.\label{figNMR}}
\end{figure}

The inset in Figure~\ref{figNMR} shows the NMR spectra measured at different temperatures
in 14~kOe field applied along $a^\ast$ direction. At this orientation, only 5 resonance
lines are resolved due to overlapping of some of the peaks. To be more specific, the
leftmost and the two rightmost peaks (at $T\geq20\,$K) are from the individual hydrogen
sites, while the central peak and the one next to the left are, respectively, three and
two superimposed individual peaks.

As can be seen in the inset in Fig.~\ref{figNMR}, the $^1$H spectrum maintains its shape
down to $\sim$20~K. At lower temperature the peaks broaden rapidly, which is most
noticeable on the right-hand side of the spectrum. In fact, below 10~K the right-hand
side becomes a featureless pedestal spreading to several MHz up in frequency. The main
panel of Fig.~\ref{figNMR} shows the temperature dependence of the leftmost peak width
measured at half height, $\Delta_{1/2}$. Rather flat above 25~K, $\Delta_{1/2}$
increases sharply below this temperature.

One of the sources for the line broadening in the $H\parallel a^\ast$ experiment could be
a \textit{b}-axis component of the internal field, $H_b$,\cite{NMR} which possibly exists
if $a^\ast$ is not a principal magnetic axis of the anisotropic material. However, the
estimate of $H_b$  using the measured dc magnetization, $M_a=9.65\,$G, for $H=14\,$kOe at
4.2~K, and the \textit{c}-axis magnetic torque measurements addressed in the next
section, give $H_b<0.2\,$G which is negligible.

Therefore the increase of the linewidth below 25~K is due to enhancement of the scatter
of the static local field induced by the electronic spin of Mn$^{2+}$ at hydrogen sites.
Assuming that the crystal structure is intact, this points to the appearance of
manganese sites with different projections of the magnetic moment on the external field
direction, i.e. magnetically inequivalent sites.
One can roughly estimate that a $\sim\,$35\% scatter of the projection of Mn$^{2+}$ moment
on the external field leads to the line broadening consistent with that
observed in the experiment at 4\,K (see Fig.~\ref{figNMR}).

In the well-known case of a commensurate N\'{e}el order, two magnetically inequivalent
sites (magnetic sublattices) result in splitting of NMR peaks,\cite{KanodaPRL75} while in
the current study a broader peak is observed instead, Fig.~\ref{figNMR}. Therefore, the
Mn$^{2+}$ spins in this system freeze below $T_{MI}$ into a more intricate magnetic state
or an incommensurate long-range order. This is probably because manganese forms a nearly
triangular network in the anion layer. In such systems with AF coupling where the
minimization of pairwise interactions is geometrically frustrated, exotic magnetic
structures are often resolved in the ground state.\cite{SpinIce}

\subsection{Magnetic torque}

\begin{figure}[h]
\includegraphics[width=0.65\linewidth,angle=90]{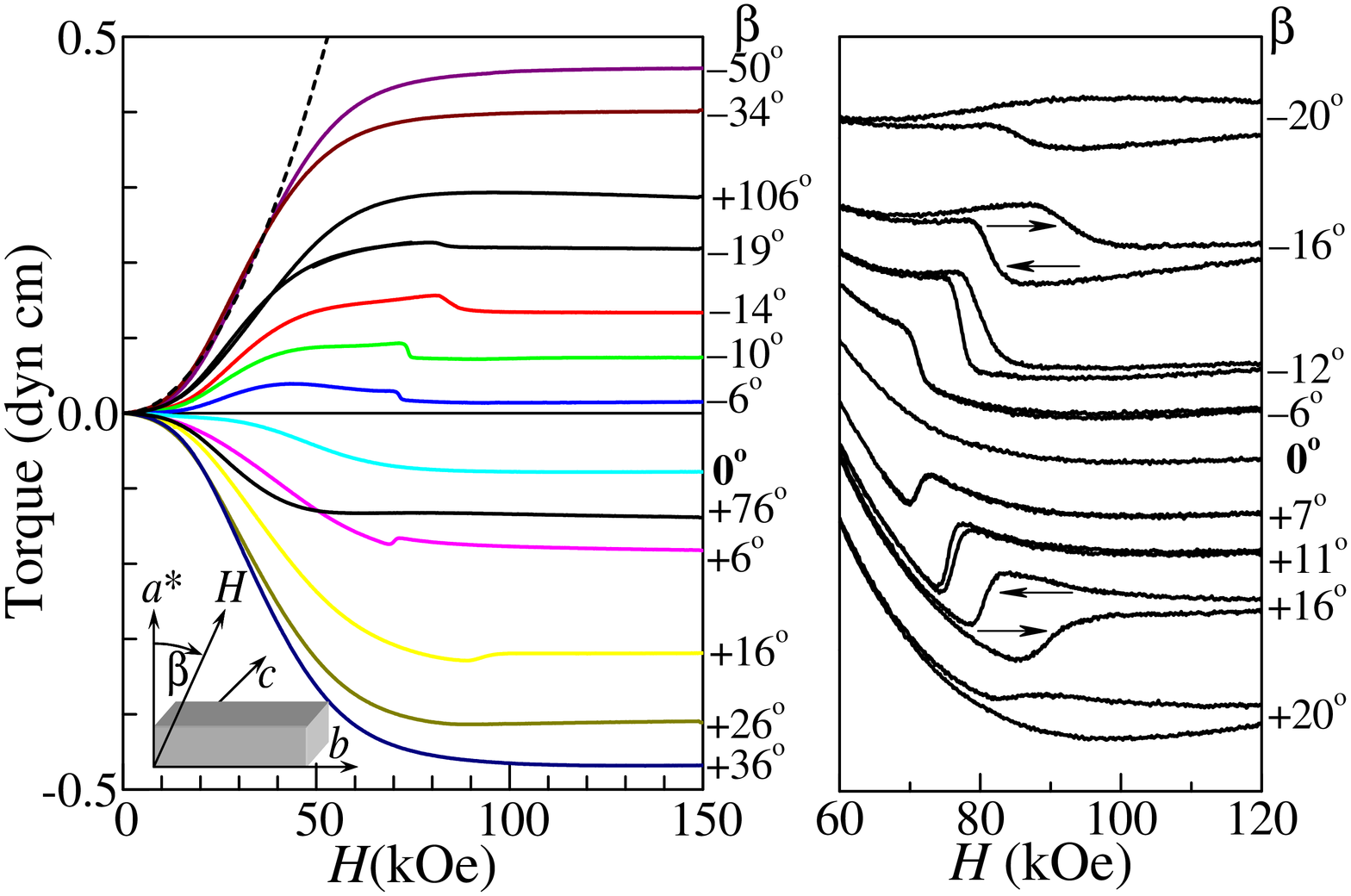}
\caption{Left panel: Field dependences of the $c$-axis torque at 1.4~K at different
angles. Dashed line: $\frac{1}{2}\Delta\chi H^2$ with $\Delta\chi=0.012\,$cm$^3$/mol, see
text. Bottom left diagram: experimental geometry. Right panel: The region with the kinks
close-up. The curves are offset along the vertical axis for better visualization. Arrows
indicate the field sweep directions. Numbers to the right of each panel indicate the
angle between the field direction and the $a^\ast$-axis. \label{figTauH}}
\end{figure}

The magnetic torque defined as $\vec{\tau}=\vec{M}\times\vec{H}$, was measured in field
sweeps between 0 and 150~kOe. The field orientation with respect to the $a^\ast$
direction, determined by polar angle $\beta$, was spanned around four directions included
in the $(bc)$-plane: $b$-axis, $c$-axis, $[0\bar{1}1]$ direction, and the direction
perpendicular to $[0\bar{1}1]$. The experimental geometry for the $c$-axis rotation is
illustrated in the bottom left of Figure~\ref{figTauH}. The torque meter is constructed
to pick up the torque component along the rotation axis because its cantilever is mounted
perpendicular to the goniometer axis.

The left panel in Figure~\ref{figTauH} shows several $c$-axis torque curves measured at
1.4~K. One can see that the main contribution to the torque monotonically develops with
the field and saturates above 100~kOe. The flattening of the torque results apparently
from the competition between the magnetic energy, which forces the saturated magnetic
moment to align with the field direction, and the magnetic anisotropy, which tends to
bring it along the axis of easy magnetization. Such behavior of the torque at high
fields/low temperatures was observed, for instance, in $\lambda$ and $\kappa$ phases of
(BETS)$_2$FeCl$_4$.\cite{taul-FeCl4, taul-FeCl4-2}

The behavior of the monotonic contribution to the torque has a $\sin2\beta$ periodicity
and is qualitatively the same for all four directions of the rotation axis. The angular
dependences of the torque at $T=1.4\,$K, $H=150\,$kOe fit the relation
$\tau(\beta)=A\sin2(\beta-\beta_0)$ with parameters $A$ and $\beta_0$ listed in
Table~\ref{TableTau}.

\begin{table}[h]
\caption{\label{TableTau} Parameters of the torque angular dependence at $T=1.4\,$K,
$H=150\,$kOe. $\beta$ is the angle between the field direction and $a^\ast$.}

\begin{ruledtabular}
\begin{tabular}{lcc}
&\multicolumn{2}{c}{$\tau(\beta)=A\sin2(\beta-\beta_0)$}\\
Rotation axis&$A$&$\beta_0(\pm1^\circ)$\\
\hline
$\parallel[0\bar{1}0]$ (\textit{b}-axis) & 3.5 & $-69^\circ$\\
$\parallel[001]$ (\textit{c}-axis) & 0.92 & $85^\circ$\\
$\parallel[0\bar{1}1]$ & 2.1 & $-66^\circ$\\
$\perp[0\bar{1}1]$ within $(bc)$ plane & 2.8 & $67^\circ$\\
\end{tabular}
\end{ruledtabular}
\end{table}

In order to understand the extent of the magnetic anisotropy given by the measured
torque, an estimate calibration of the torque meter was done using the cantilever
geometry and the Young's modulus of its material. Calculations give a factor
$\sim 0.5\,$dyn$\cdot$cm/unit for the vertical scale in Fig.~\ref{figTauH} and
for the $A$ parameter in Table~\ref{TableTau}.

As one can see in Fig.~\ref{figTauH}, at low fields the torque is approximately quadratic
in field. This behavior expressed as $\tau=\frac{1}{2}\Delta\chi H^2\sin2(\beta+\beta_0)$
can be used to estimate the susceptibility anisotropy, $\Delta\chi$, in the plane normal
to the rotation axis. Using the above calibration factor, one obtains $\Delta\chi_{a^\ast
b}\simeq0.012\,$cm$^3$/mol for the \textit{c}-axis torque. In a similar way, for the
\textit{b}-axis torque our data gives $\Delta\chi_{a^\ast c}\simeq0.03\,$cm$^3$/mol. The
anisotropy is therefore $\leq5\%$ of the susceptibility (0.66~cm$^3$/mol at 2~K), which
appears negligibly small in the dc magnetization measurements presented above. However,
these values are at least an order of magnitude bigger compared to organic compounds with
non-magnetic anion layers.\cite{KanodaPRL75, WernerJPF114} Therefore the dominating
contribution to the torque comes from the anisotropy in the Mn$^{2+}$ magnetic subsystem.
Importantly, none of the principal axes of this magnetic subsystem coincide with the
crystallographic axes (see Table \ref{TableTau}) because the torque along a principal
magnetic axis is identically zero. Moreover, well above the MI transition the torque
still zeroes at angles close to $\beta_0$  which means that the orientation of the
principal axes of Mn$^{2+}$ subsystem does not change appreciably at $T_{MI}$.

It should be mentioned that while the high-field torque zeroes at angles $\beta=\beta_0$
and $\beta_0\pm\pi/2$, at low temperature the low-field torque ($H<50$\,kOe) for these
directions does not vanish completely showing a small bell-like feature, as has also been
noticed in Ref.~\onlinecite{r12}. The origin of this feature is so far unclear and
requires further investigation. Anyway, the periodicity of this feature is a regular
$\sin2\beta$. Therefore, it cannot be ascribed to a conventional spin-flop transition in
a uniaxial antiferromagnet, which is periodic as
$\rm{sgn}\beta\cos\beta$.\cite{YoshidaSF}

As one can see in Fig.~\ref{figTauH}, at small angles $\beta$ the monotonic development
of the torque is broken by a step-like feature, or a ``kink". A similar behavior, though
with somewhat smaller kinks, has also been found for the field rotations around
$[0\bar{1}1]$ and around the axis perpendicular to $[0\bar{1}1]$, whereas no kinks have
been detected in the $b$-axis torque curves.

\begin{figure}[h]
\includegraphics[width=0.65\linewidth,angle=90]{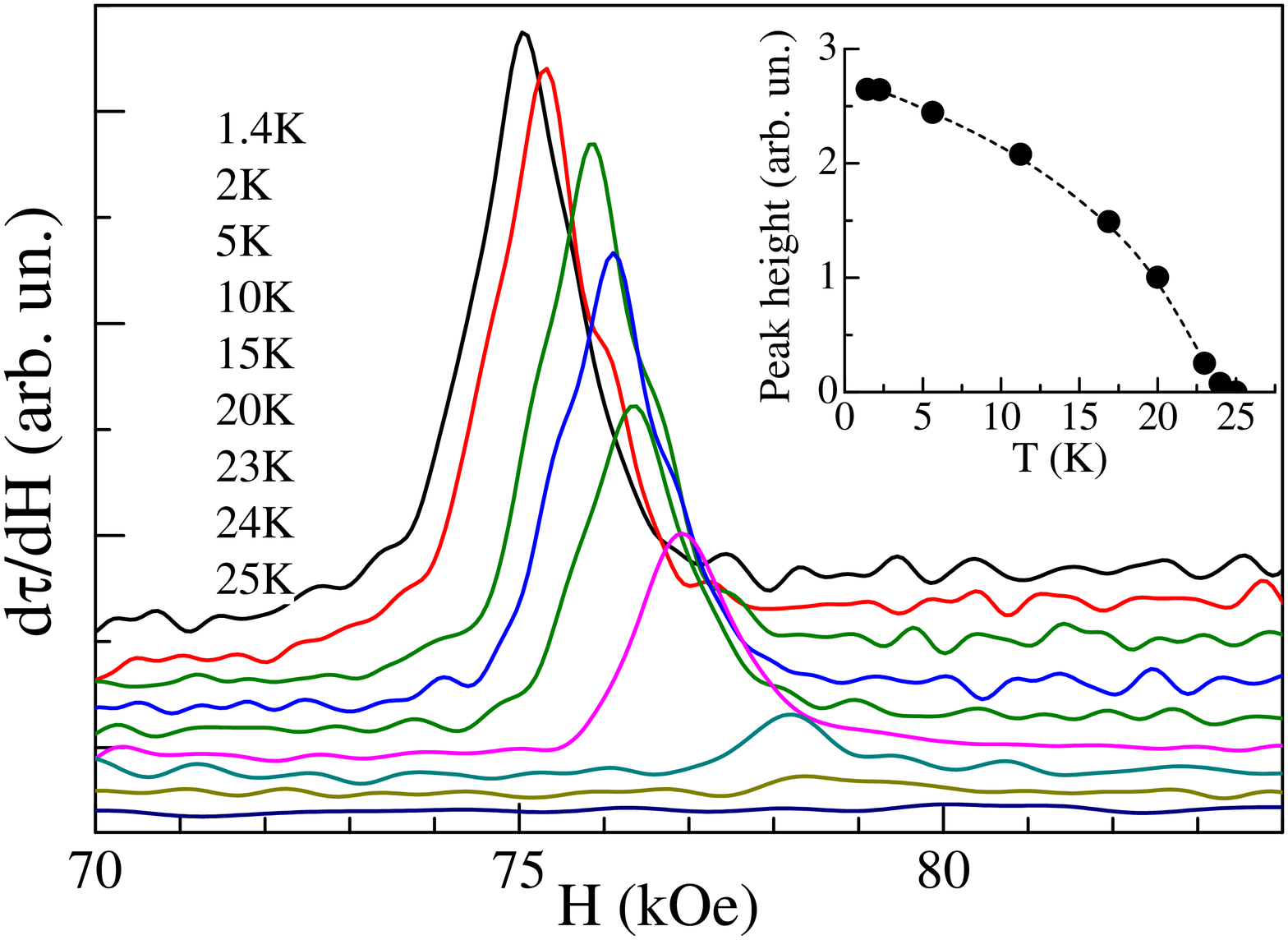}
\caption{(Color online) The field derivative of the torque, $d\tau/dH$, as a function of
field for $\beta=+10^\circ$, at temperatures from 1.4\,K (the top curve) to 25\,K (the
bottom curve). Inset: Temperature dependence of the $d\tau/dH$ peak
height.\label{figKinksT}}
\end{figure}

The angular evolution of the kink is shown in more detail in the right panel in
Fig.~\ref{figTauH}. The behavior of the kink with the field and the angle is quite
peculiar. First, it emerges in a rather symmetric way with tilting the field from
$H\parallel a^\ast$ ($\beta=0$ and $\pm\pi$) and is not seen around $H\perp a^\ast$
($\beta=\pm\pi/2$): compare curves for $\beta=-14^\circ$ and $+76^\circ$, and for
$\beta=+16^\circ$ and $+106^\circ$ in the right panel in Fig.~\ref{figTauH}. The kink
periodicity is therefore $\rm{sgn}\beta\cos\beta$, which is the same as for the torque at
the spin-flop transition. Secondly, the kink moves up in field with increasing the tilt
from the $a^\ast$ direction, and vanishes when the tilt is more than 20$^\circ$, which
means that we deal with a projection of some preferential direction on $a^\ast$. Thirdly,
it becomes increasingly hysteretic at $\beta>\pm 10^\circ$ indicating a first-order
transition. With increasing temperature, the kink becomes smaller and disappears above
25~K, as illustrated in Figure~\ref{figKinksT} where the $c$-axis torque field
derivative, $d\tau/dH$, for $\beta=+10^\circ$ is plotted for temperatures from 1.4 to
25~K. The inset in Fig.~\ref{figKinksT} shows temperature dependence of the $d\tau/dH$
peak height. The listed peculiarities of the kink behavior in the insulating state
indicate an abrupt field-induced reorientation of magnetic moments, quite reminiscent of
a spin-flop transition in an antiferromagnet.

The direction $a^\ast$ in which vicinity the spin-flop occurs should be close to the AF
easy axis. As noted above, the $a^\ast$ direction is not the principal axis of the
magnetic subsystem associated with Mn$^{2+}$ network. We therefore speculate that the
kink is related to another magnetic subsystem, namely to localized $\pi$-electron spins
in BETS layers. Indeed, if the MI transition in the present compound is caused by the
Mott instability,\cite{ZverevPT} one can expect the spins associated with the localized
$\pi$ electrons to be antiferromagnetically ordered. While we have currently no
information about the principal axes of the $\pi$-electron spin system, it is likely that
one of them is along $a^{\ast}\perp bc$, which is of course a special direction in this
layered electronic system. The size of the kink, $\sim 10$ times smaller than the maximum
torque in Fig. 4, is also consistent with what one would expect as a contribution from
$\pi$ electrons on BETS molecules.\cite{KanodaPRL75, WernerJPF114}

There is a number of features that distinguish the observed field-induced transition from
the usual spin-flop in a uniaxial antiferromagnet. In the latter case the spin-flop
feature in the torque has a sharp $\lambda$-like shape. Such spin-flop transition occurs
in $\lambda$-(BETS)$_2$FeCl$_4$ \cite{KobaPolyhedron} but at much lower field 12\,kOe.
However, in $\lambda$-(BETS)$_2$FeCl$_4$ the Fe$^{3+}$-based anion subsystem is deeply
involved into the formation of the AF order. This is inferred from the large drop of the
static susceptibility \cite{BrossFe} and the huge spin-flop feature in the torque
\cite{KobaPolyhedron} comparable with the maximum high-field torque. On the contrary, in
the compound under study a long-range ordering in the Mn$^{2+}$ magnetic subsystem is not
as obvious. The geometrically frustrated Kagome-type lattice of Mn$^{2+}$ is reluctant of
N\'{e}el-type ordering, as can be concluded from the minor changes of the susceptibility
(see Fig.~\ref{figMvsT}). In turn, the frustrated Mn$^{2+}$ subsystem can essentially
modify the AF structure of the BETS $\pi$-electrons due to the $\pi-d$ interactions.

\section{Conclusion}

We have performed dc magnetization, $^1$H NMR and magnetic torque studies of organic
conductor $\kappa$-(BETS)$_2$Mn[N(CN)$_2$]$_3$ which undergoes a MI transition at 25~K.
The magnetization above $T_{MI}$ follows the Curie-Weiss law, revealing antiferromagnetic
interactions between $S=5/2,\,L=0$ Mn$^{2+}$ spins, and violates it at lower temperature.
The $^1$H NMR spectrum determined by dipolar fields from Mn$^{2+}$ moments, exhibits vast
broadening of the resonance peaks below 25~K resulting from freezing of Mn$^{2+}$ spins.
Both the magnetization and NMR data obtained at fields below 70~kOe, therefore suggest a
transition from paramagnetic to a frozen state in the manganese subsystem taking place at
$T_{MI}$. However, the resulting state clearly differs from the conventional commensurate
N\'{e}el type antiferromagnetic order.

Yet another signature of a magnetic transition is the kink in the field-dependent torque
emerging right below $T_{MI}$. This one takes place above $\sim70\,$kOe and possesses
many properties of a spin-flop transition. Most likely it occurs within the spins of
Mott-localized $\pi$ electrons of BETS molecules. If so, it would be more natural to
consider the Mott localization as a driving force for both the MI transition and magnetic
transformations in the system. An additional argument towards this conclusion is a low
$\theta\sim -5\,$K in the Curie-Weiss temperature dependence of the spin susceptibility
(compared to $T_{MI}\approx25\,$K): in a conventional antiferromagnet the AF transition
usually occurs at temperature lower than $|\theta|$.\cite{Morrish} To further clarify
this issue, NMR studies at higher fields where the kink appears would be very helpful.

\section{Acknowledgements}

This work was supported by the RFBR grants 07-02-91562, 10-02-01202 and DFG grant
RUS~113/926/0. The authors gratefully acknowledge the assistance in the questions of
crystallography from R.~P.~Shibaeva, S.~S.~Khasanov and S.~V.~Simonov and the technical
support from N.~A.~Belov.


\begin{thebibliography}{99}

\bibitem{ishi88}
T.~Ishiguro, K.~Yamaji, and G.~Saito, {\it Organic superconductors} (Springer-Verlag,
Berlin, 1988).

\bibitem{lebe08}
{\it The Physics of Organic Superconductors and Conductors}, edited by A.~G.~Lebed
(Springer Berlin Heidelberg, 2008).

\bibitem{r6-1} E.~Coronado and P.~Day, Chem.~Rev. \textbf{104}, 5419 (2004); T.~Enoki and
A.~Miyazaki, \textit{ibid.} \textbf{104}, 5449 (2004); H.~Kobayashi, H.~Cui, and
A.~Kobayashi, \textit{ibid.} \textbf{104}, 5265 (2004).

\bibitem{r6-4} L.~Ouahab, in {\it Organic Conductors, Superconductors and Magnets: From Synthesis to Molecular
Electronics} edited by L.~Ouahab and E.~Yagubskii (Kluwer~Acad.~Publ.
Dordrecht/Boston/London, 2003), p.~99.

\bibitem{r6-5} G.~Saito and Y.~Yoshida, Bull.~Chem.~Soc.~Jpn. \textbf{80}, 1 (2007).

\bibitem{r6-6} E.~Coronado and K.~R.~Dunbar, Inorg.~Chem. \textbf{48}, 3293 (2009).

\bibitem{r6-7} L.~Ouahab and T. ~Enoki, Eur.~J.~Inorg.~Chem. \textbf{2004}, 933 (2004).

\bibitem{r10} S.~Uji, H.~Shinagawa, T.~Terashima, T.~Yakabe, Y.~Terai, M.~Tokumoto,
A.~Kobayashi, H.~Tanaka, and H.~Kobayashi, Nature \textbf{410}, 908 (2001).

\bibitem{r11} H.~Fujiwara, H.~Kobayashi, E.~Fujiwara, and A.~Kobayashi, J.~Am.~Chem.~Soc.
\textbf{124}, 6816 (2002).

\bibitem{r12} N.~D.~Kushch, E.~B.~Yagubskii, M.~V.~Kartsovnik, L.~I.~Buravov, A.~D.~Dubrovskii,
A.~N.~Chekhlov, and W.~Biberacher, J.~Am.~Chem.~Soc. \textbf{130}, 7238 (2008).

\bibitem{ZverevPT} V.~N.~Zverev, M.~V.~Kartsovnik, W.~Biberacher,
S.~S.~Khasanov, R.~P.~Shibaeva, L.~Ouahab, L.~Toupet, N.~D.~Kushch, E.~B.~Yagubskii, and
E.~Canadell, Phys.~Rev.~B \textbf{82}, 155123 (2010).

\bibitem{BrossFe} L.~Brossard \textit{et al.}, Eur.~Phys.~J.~B \textbf{1}, 439 (1998).

\bibitem{hott00} C.~Hotta and H.~Fukuyama, J.~Phys.~Soc.~Jpn. {\bf 69}, 2577 (2000).

\bibitem{AkibaSH} H.~Akiba, S.~Nakano, Y.~Nishio, K.~Kajita, B.~Zhou,
A.~Kobayashi, and H.~Kobayashi, J.~Phys.~Soc.~Japan \textbf{78}, 033601 (2009).

\bibitem{BrooksMoess} J.~C.~Waerenborgh, S.~Raba\c{c}a, M.~Almeida, E.~B.~Lopes, A.~Kobayashi, B.~Zhou, and
J.~S.~Brooks, Phys.~Rev.~B \textbf{81}, 060413(R) (2010).

\bibitem{chri94}  P. Christ, W. Biberacher, H. M\"{u}ller, K. Andres,
Solid State Commun. {\bf 91}, 451 (1994).

\bibitem{NMR} For the field in the (\textit{ac}) plane, every of the 8 hydrogen sites of
the BETS molecule is mapped to a single line in NMR spectrum. However, a \textit{b}-axis
field component doubles the number of magnetically inequivalent hydrogen sites (hence the
number of NMR peaks) due to the mirror-type $(x,y,z)\rightarrow (x,-y,z)$ symmetry
operations of the space group $P$2$_1$/$c$ (2-fold screw axis with direction [010] and
glide plane normal to [010]).

\bibitem{KanodaPRL75} K.~Miyagawa, A.~Kawamoto, Y.~Nakazawa, and K.~Kanoda, Phys.~Rev.~Lett. \textbf{75}, 1174
(1995).

\bibitem{SpinIce} S.~T.~Bramwell and M.~J.~P.~Gingras, Science \textbf{294}, 1495 (2001).

\bibitem{taul-FeCl4} T.~Sasaki, H.~Uozaki, S.~Endo, and N.~Toyota, Synth.~Metals \textbf{120}, 759 (2001).

\bibitem{taul-FeCl4-2} H.~Uozaki, T.~Sasaki, S.~Endo, and N.~Toyota, J.~Phys.~Soc.~Jpn. \textbf{69}, 2759 (2000).

\bibitem{WernerJPF114} W.~Biberacher, P.~Christ, M.~V.~Kartsovnik, D.~Andres, H.~M\"{u}ller, and
N.~Kushch, J.~Phys.~IV~France \textbf{114}, 291 (2004).

\bibitem{YoshidaSF} K.~Yoshida, Prog.~Theor.~Phys. \textbf{6}, 691 (1951).

\bibitem{KobaPolyhedron} M.~Tokumoto, H.~Tanaka, T.~Otsuka, H.~Kobayashi, and
A.~Kobayashi, Polyhedron \textbf{24}, 2793 (2005).

\bibitem{Morrish} A.~H.~Morrish, \textit{The Physical Principles of Magnetism} (Wiley-IEEE
Press, New York, 2001).

\end{thebibliography}
\end{document}